\documentclass[12pt]{article}

\usepackage{scicite}
\usepackage{amsmath}
\usepackage{times}
\usepackage[pdftex]{graphicx}
\usepackage[figuresright]{rotating}%
\usepackage{amsmath,amssymb,amsfonts}
\usepackage{textcomp}
\usepackage{xcolor}
\usepackage{braket}
\usepackage{multirow}
\usepackage{amsthm}
\usepackage{algorithm}
\usepackage{algorithmic}

\def\citepunct{,}
\def\citedash{--}
\newcommand{\system}{QiankunNet } 
\newcommand{\fesa}{[2Fe-2S]}
\newcommand{\fesb}{[4Fe-4S]}

\newcommand{\boldx}{\bold{x}}

\newenvironment{sciabstract}{%
\begin{quote} \bf}
{\end{quote}}

\title{Transformer-Based Neural Networks Backflow for Strongly Correlated Electronic Structure}
 
\author
{Huan Ma$^{1,2,\dagger}$,  Bowen Kan$^{3,\dagger}$, Honghui Shang$^{1,2\ast}$, Jinlong Yang$^{1,2\ast}$\\
\\
\normalsize{$^{1}$ State Key Laboratory of Precision and Intelligent Chemistry,} \\ \normalsize{University of Science and Technology of China, Hefei 230026, China}\\
\normalsize{$^{2}$ Hefei National Laboratory,} \\   \normalsize{University of Science and Technology of China, Hefei 230088, China}\\
\normalsize{$^{3}$  Institute of Computing Technology, Chinese
 Academy of Sciences, Beijing 100190, China} 
\\
\normalsize{$^\dagger$These authors contribute equally to this work.}\\
\normalsize{$^\ast$To whom correspondence should be addressed:  shanghui.ustc@gmail.com;  jlyang@ustc.edu.cn.}\\
}

\date{}

\begin{document} 


\baselineskip24pt


\maketitle


\begin{sciabstract}
Solving the electronic Schrödinger equation for strongly correlated systems remains one of the grand challenges in quantum chemistry.  Here we demonstrate that Transformer architectures can be adapted to capture the complex grammar of electronic correlations through neural network backflow. In this approach, electronic configurations are processed as token sequences, where attention layers learn non-local orbital correlations and token-specific neural networks map these contextual representations into backflowed orbitals.  Application to strongly correlated iron-sulfur clusters validates our approach: for $\left[\mathrm{Fe}_2 \mathrm{~S}_2\left(\mathrm{SCH}_3\right)_4\right]^{2-}$ (\fesa) (30e,20o), the ground-state energy within chemical accuracy of DMRG  while predicting magnetic exchange coupling constants closer to experimental values than all compared methods including DMRG, CCSD(T), and recent neural network approaches. For $\left[\mathrm{Fe}_4 \mathrm{S}_4\left(\mathrm{SCH}_3\right)_4\right]^{2-}$ (\fesb) (54e,36o), we match DMRG energies and accurately reproduce detailed spin-spin correlation patterns between all Fe centers. The approach scales favorably to large active spaces inaccessible to exact methods, with distributed VMC optimization enabling stable convergence. These results establish Transformer-based backflow as a powerful variational ansatz for strongly correlated electronic structure, achieving superior magnetic property predictions while maintaining chemical accuracy in total energies.
\end{sciabstract}

\section*{Introduction}
Accurately solving the electronic Schrödinger equation for complex, strongly correlated systems remains a grand challenge in chemistry and physics. The full configuration-interaction (FCI) method provides exact solutions in principle, but its exponential cost restricts it to small systems (on the order of 26 electrons in 23 orbitals at most
)~\cite{Gao2024}. Practical electronic structure methods therefore rely on approximations: mean-field approaches like Hartree–Fock and density functional theory (DFT) scale to large molecules but often break down when electron correlation is strong, while post-Hartree–Fock techniques ~\cite{Perturbation,MP2,White1992,White1993,McMillan1965,FoulkesRajagopal2001,AustinLester2012} such as truncated configuration interaction~\cite{MCSCF} and coupled-cluster~\cite{Bartlett2007} capture many correlations but can still fail for strongly correlated systems. A prototypical example is the iron-sulfur clusters, which are notorious for their dense manifold of near-degenerate spin states and severe electron correlation and standard methods struggle to predict even the qualitative magnetic ground state due to intricate electron  interactions. This failure underscores a broader need for new wavefunction ansätze that can simultaneously achieve high accuracy in total energies and correctly capture complex properties such as spin couplings.

In recent years, neural network quantum states (NNQS) provide a new pathway by directly parameterizing the many-body wavefunction with machine learning models~\cite{CarleoTroyer2017}. By leveraging the universal approximation capabilities of neural networks, NNQS can, in principle, efficiently represent wavefunctions in exponentially large Hilbert spaces while maintaining polynomial scaling for optimization. Subsequent developments have diversified along two trajectories: first quantization~\cite{PfauFoulkes2020,HermannNoe2020,vonglehn2022psiformer}, which works in continuous real space  and second quantization~\cite{ChooCarleo2020,BarrettLvovsky2022,Zhao2023,PRL2023,Zhang2023,Liu2024-1,Wu2023,Fu2024,Kan2025}, which operates in discrete orbital bases. 
Across both paradigms, the most challenging strongly correlated systems—such as iron–sulfur clusters—have remained problematic.

The Transformer architecture has emerged as one of the most influential paradigms in modern machine learning with exceptional versatility, scalability, and generalization ability~\cite{transformer2017,radford2018gpt,radford2019gpt2,Brown2020}.  More recently, Transformer-based wavefunction models have begun to show promise in quantum science~\cite{Zhang2023,PRL2023,Glehn2022, Wu2023}, where its capacity to capture long-range dependencies resonates with the highly entangled nature of many-body quantum states. Despite this progress, the systematic deployment of Transformer-based architectures for solving electronic Schrödinger equations in complex molecular systems remains underexplored. The central open question is whether such models can be enhanced to tackle the strong correlated systems—like \fesb and other multi-metal clusters—capturing not only ground-state energies but also the delicate magnetic couplings and multi-reference character that define their chemistry? Here we show that the answer is yes – by integrating a classic insight from many-body physics, the backflow correlation\cite{Feynman1956}, into the heart of a transformer-based NNQS. Building on this idea, we introduce a backflow-enhanced extension of QiankunNet. This framework harnesses the expressive power of Transformer attention mechanisms to parameterize configuration-dependent orbitals. By embedding backflow transformations into a Transformer architecture, \system achieves a unique combination of physical fidelity, expressive flexibility, and computational scalability. We benchmark our approach on paradigmatic strongly correlated transition-metal clusters, including the prototypical \fesa\ and \fesb\ complexes central to bioinorganic chemistry. These systems present a notorious challenge for conventional methods due to their dense manifold of spin states and severe static correlation. \system achieves chemical accuracy in describing the relative energetics of these clusters, and demonstrating scalability to active spaces far beyond the reach of exact diagonalization. By uniting the Transformer architecture with neural network backflow, \system establishes a new foundation for solving the molecular Schrödinger equation in strongly correlated regimes.


\begin{figure}
    \centering    \includegraphics[width=1.0\linewidth]{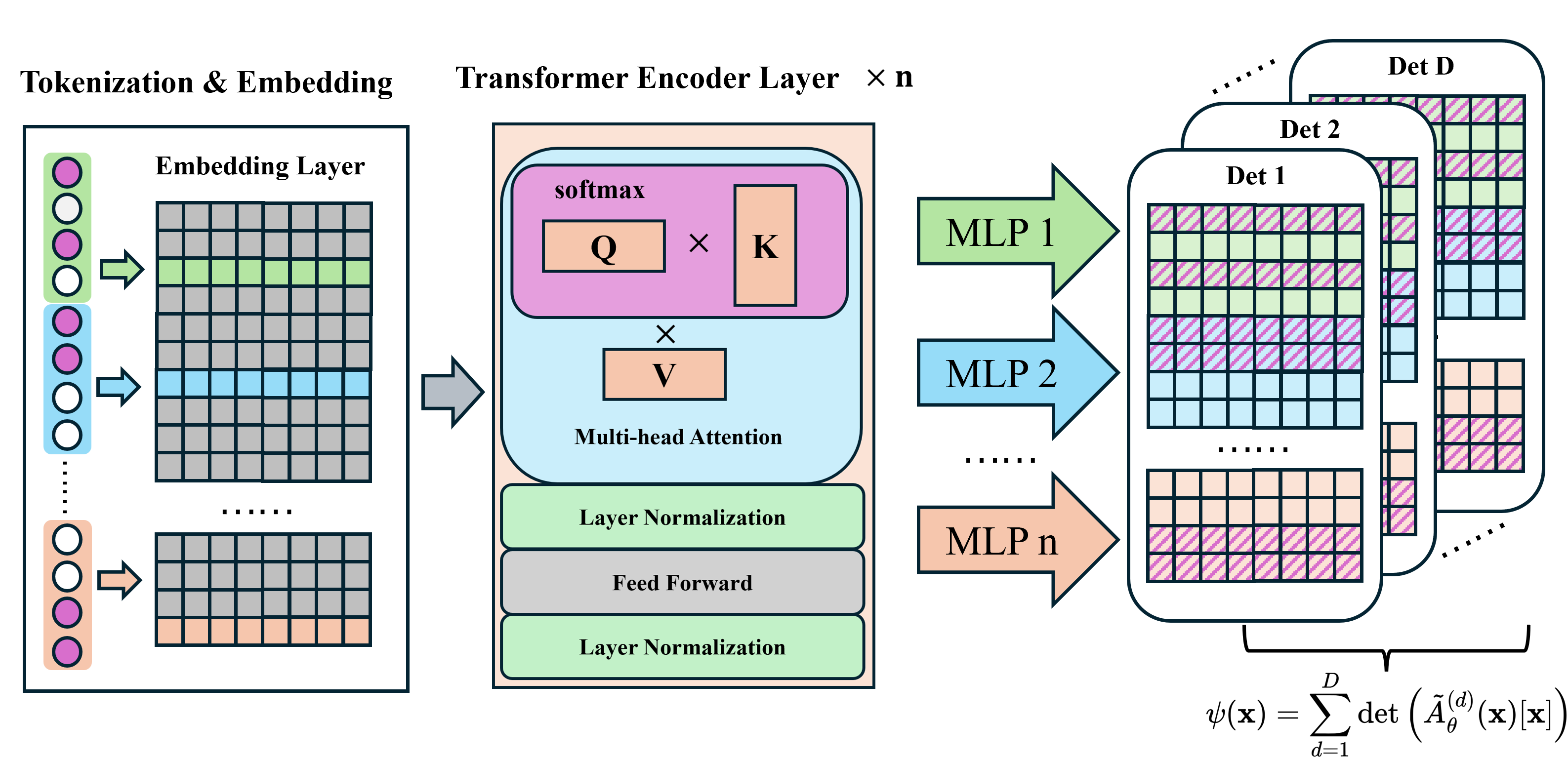}
\caption{Schematic computational workflow of \system. The procedure begins with a binary electron configuration string segmented into tokens representing local spin-orbital groups. Each token is embedded into a feature vector via a learnable lookup table (embedding layer), then processed through Transformer encoder layers to capture long-range electron correlations. The resulting features are mapped by token-specific MLPs to construct the single-particle orbital (SPO) matrix blocks, which are assembled into $D$ distinct SPO matrices $\tilde{A}_{\theta}^{(d)}(\mathbf{x})$. For each configuration $\mathbf{x}$, occupied orbitals are selected to form $N_e \times N_e$ matrices, and their determinants are summed to yield the final wavefunction $\psi(\mathbf{x})$.}
\end{figure}

\section*{Framework Overview}
\label{sec:framework}
A central concept underpinning our approach is the backflow transformation, which dynamically re-defines the single-particle orbital basis as a function of the many-electron configuration. Rather than using a fixed set of molecular orbitals for all configurations (as in a conventional Slater determinant), backflow allows each electron’s effective orbital to “flow” in response to the positions or occupancies of all the others. This idea – dating back to Feynman and Cohen’s work on correlated fluids~\cite{Feynman1956} – introduces nonlinearity and correlation inside the determinant, enabling the wavefunction to adapt its sign and amplitude to electron interactions. Recent neural-network wavefunction studies have embraced backflow~\cite{Liu2024-1,liu2025NNBF}  as a powerful way to encode correlation beyond the reach of cluster expansions or Jastrow factors. By making the orbital basis configuration-dependent, one can directly capture entangled multi-electron patterns in a compact ansatz. In short, backflow endows the wavefunction with context-aware orbitals, which is especially crucial for strongly correlated systems where a single reference configuration is inadequate.

We implement these ideas in \system\ through a Transformer-based neural backflow architecture. For a molecular system containing $N_e$ electrons, a basis set composed of $N_o(> N_e)$ spin orbitals
$\mathcal{B}=\{\lvert \phi_i\rangle\}_{i=1}^{N_o}$ is given to define a many-electron wavefunction in the form
\begin{equation}
    \lvert \psi \rangle \;=\; \sum_{i}\, \psi(\mathbf{x}_i)\,\lvert \mathbf{x}_i \rangle,
\end{equation}
where $\lvert \mathbf{x}_i \rangle = \lvert x^{(i)}_{1},\ldots,x^{(i)}_{N_o}\rangle$ is the $i$th computational basis vector in second
quantization. Here, $x^{(i)}_{j}\in\{0,1\}$ denotes whether the $j$th spin orbital is occupied in the $i$th
computational basis vector. In \system, the many-electron wavefunction is parameterized by a Transformer that outputs configuration-dependent single-particle orbitals (SPOs), which are then fed into a Slater determinant. For a given electronic configuration $\mathbf{x}$ (specified by occupation of a set of spin-orbitals), the ansatz reads
\begin{equation}
\psi(\mathbf{x}) \;=\; \det\!\big(\tilde{A}_{\theta}(\mathbf{x})[\mathbf{x}]\big)\,,
\end{equation}
where $\tilde{A}_{\theta}(\mathbf{x})\in\mathbb{R}^{N_{\mathrm{so}}\times N_e}$ is the orbital matrix generated by the neural network ($N_{\mathrm{so}}$ spin-orbitals, $N_e$ electrons), and $\tilde{A}_{\theta}(\mathbf{x})[\mathbf{x}]\in\mathbb{R}^{N_e\times N_e}$ denotes the sub-matrix formed by the rows corresponding to the occupied spin-orbitals in $\mathbf{x}$. Thus, the network produces SPOs \emph{on the fly} for each configuration, and the wavefunction amplitude is given by the determinant restricted to the occupied positions. If multiple Slater determinants $D$ are used, the network can output $D$ independent orbital matrices which are summed in the final wavefunction—analogous to a multi-determinant expansion.  

Figure~1 illustrates the architecture of \system. To input a configuration into the Transformer, we represent electron occupancy as a length-$N_{\mathrm{so}}$ binary string. We partition this string into groups of $t$ spin-orbitals per token so that each token encodes a local occupancy pattern (one of $2^t$ possibilities). Each token is embedded via a learned lookup table into a $d_f$-dimensional feature vector. The collection of token embeddings forms the initial input matrix $X^{(0)} \in \mathbb{R}^{N_t \times d_f}$ with $N_t=\lceil N_{\mathrm{so}}/t\rceil$.  We employ a stack of $L_e$ Transformer encoder layers to process these token features. Each layer applies multi-head self-attention followed by a position-wise feed-forward transformation, with residual connections and layer normalization (cf.\ Eq.~(3) in the main text). At encoder layer $l$, queries, keys, and values are computed as
\begin{equation}
Q \;=\; X^{(l-1)} W_Q^{(l)}, \qquad
K \;=\; X^{(l-1)} W_K^{(l)}, \qquad
V \;=\; X^{(l-1)} W_V^{(l)}\,,
\end{equation}
and the multi-head attention output is
\begin{equation}
\mathrm{MHA}^{(l)}(Q,K,V) \;=\;
\sum_{h=1}^{N_h}
\operatorname{softmax}\!\left(\frac{Q_h K_h^{\top}}{\sqrt{d_{\text{atten}}/N_h}}\right) V_h\,,
\end{equation}
with $N_h$ attention heads (each head uses its own projections $Q_h,K_h,V_h$; $d_{\text{atten}}$ denotes the attention hidden dimension). Following attention, a feed-forward network applies a nonlinear transformation to each token:
\begin{equation}
\mathrm{FFN}^{(l)}(H) \;=\; \operatorname{ReLU}\!\left(H\, W_{\mathrm{FFN}}^{(l)} + b_{\mathrm{FFN}}^{(l)}\right),
\end{equation}
where $H$ is the output of the attention sub-layer. Stacking these layers yields an output matrix $\tilde{X}\in\mathbb{R}^{N_t\times d_f}$ that encodes high-order correlations across tokens.  A set of small multi-layer perceptrons (MLPs) transforms each token’s final feature vector $\tilde{X}_i$ into a block of $t$ rows of the SPO matrix. For token $i$, an MLP outputs a sub-matrix $\tilde{A}_{\theta,i}(\mathbf{x})\in\mathbb{R}^{t\times N_e}$; stacking over tokens forms the full $\tilde{A}_{\theta}(\mathbf{x})$. Each token-specific MLP has $L_m$ hidden layers and, if multiple determinants are used, produces $D$ distinct SPO blocks per token. The MLPs share architecture but have independent parameters per token. A final linear layer outputs the orbital values for that token. Assembling the token-wise outputs and selecting the occupied rows yields the Slater determinant coefficient $\det\!\big(\tilde{A}_{\theta}(\mathbf{x})[\mathbf{x}]\big)$.  To enhance expressiveness, QiankunNet can use a linear combination of $D$ configuration-dependent determinants, analogous to a multi-determinant expansion. Practically, each token-MLP’s final layer is expanded to output $D\cdot t \cdot N_e$ values, which are reshaped into $D$ sub-matrices. The wavefunction becomes
\begin{equation}
\psi(\mathbf{x}) \;=\; \sum_{d=1}^{D} \det\!\Big(\tilde{A}^{(d)}_{\theta}(\mathbf{x})[\mathbf{x}]\Big)\,,
\end{equation}
improving accuracy for strongly correlated systems at the cost of more parameters. Unless stated otherwise, we set $D=2$ in this work (see Table~2).

We optimize parameters $\theta$ by variational Monte Carlo (VMC). The variational energy is
\begin{equation}
    E_{\theta} \;=\; \frac{\langle \Psi_{\theta} | \hat{H} | \Psi_{\theta} \rangle}{\langle \Psi_{\theta} | \Psi_{\theta} \rangle}\,,
\end{equation}
estimated stochastically by sampling configurations $\mathbf{x}$ from $p(\mathbf{x})=|\psi_{\theta}(\mathbf{x})|^2/\sum_{\mathbf{x}^{\prime}}|\psi_{\theta}(\mathbf{x}^{\prime})|^2$. For any operator $\hat{O}$, the Monte Carlo estimator is $\langle \hat{O}\rangle \approx \mathbb{E}_{\mathbf{x}\sim p(\mathbf{x})}[O_{\mathrm{loc}}(\mathbf{x})]$ with local value
\begin{equation}
    O_{\mathrm{loc}}(\mathbf{x})\;=\;\frac{\langle \mathbf{x}|\hat{O}|\Psi_{\theta}\rangle}{\langle \mathbf{x}|\Psi_{\theta}\rangle}\,.
\end{equation}

In particular, the local energy (for $\hat{O}=\hat{H}$) is
\begin{equation}
E_{\mathrm{loc}}(\mathbf{x}) \;=\; \sum_{\mathbf{x}^{\prime}} H_{\mathbf{x} \mathbf{x}^{\prime}} \,\frac{\psi_{\theta}(\mathbf{x}^{\prime})}{\psi_{\theta}(\mathbf{x})}\,,
\end{equation}
$H_{\mathbf{x} \mathbf{x}^{\prime}}=\langle \mathbf{x}|\hat{H}| \mathbf{x}^{\prime} \rangle$. We employ gradient-based optimization; the energy gradient can be written in terms of fluctuations of $E_{\mathrm{loc}}$ times the covariant derivatives of $\ln \psi_{\theta}$. In practice, we combine Adam for rapid initial convergence with a second-order stochastic reconfiguration method (MARCH) for refinement. MARCH approximates natural-gradient preconditioning via an inverse Fisher-like matrix with momentum for stability; we use a distributed implementation to handle large parameter counts. During training, we periodically evaluate energies and observables (e.g., spin correlations) on independent samples to monitor convergence.

A computationally demanding aspect of our approach lies in the exact evaluation of the local energy $E_{\text{loc}}=\sum_{\mathbf{x}}H_{\mathbf{xx}^{\prime}}\frac{\psi(\mathbf{x}^{\prime})}{\psi(\mathbf{x})}$. Deterministic computation requires enumerating all configurations \(\mathbf{x}^{\prime}\) connecting with the non-zero Hamiltonian matrix element \(H_{\mathbf{x} \mathbf{x}^{\prime}}\) which scales as $O(N^4)$.  As a result, computing the wavefunction amplitudes \(\psi(\mathbf{x}^{\prime})\) constitutes the primary computational bottleneck. To address this, we employ thread-level parallelism to accelerate the generation of coupled configurations and each thread executes its own sampling process, leveraging a semi-stochastic scheme to minimize the number of expensive $\psi(\mathbf{x}^{\prime})$ evaluations.
In this scheme, configurations are distributed across multiple cores, with each thread performing coupling calculations only for its assigned configurations. The resulting coupled states and $H_{\mathbf{x} \mathbf{x}^{\prime}}$ from all threads are then reduced to thread 0. Coupling paths are classified into a deterministic set $\mathcal{D}$ and a candidate set $\mathcal{C}$. Elements in $\mathcal{C}$ are normalized and sampled by probability $P(\mathbf{x}^{\prime})=\frac{|H^{\mathbf{x}^{\prime}}_{e}|}{\sum_{x'}|H^{\mathbf{x}^{\prime}}_{e}|}$, significantly reducing the computational load. The global sets $\mathcal{D}$ and $\mathcal{C}$ are then consolidated, and the total energy expectation $E_{\text{loc}}$ is computed efficiently via vectorized operations after obtaining $\psi(\mathbf{x}^{\prime})$.

The use of a Transformer backbone endows the model with great representational power to capture complex correlation patterns (the attention mechanism learns how each orbital’s occupancy influences every other), while the backflow determinant construction imposes a meaningful inductive bias reflecting known quantum-chemical structure. Unlike black-box function approximators, \system’s outputs (the backflow orbitals) have direct physical interpretation, potentially allowing insight into the emergent effective orbitals and correlation motifs it learns. 

\begin{figure}[ht]
    \centering
    \includegraphics[width=1.0\linewidth]{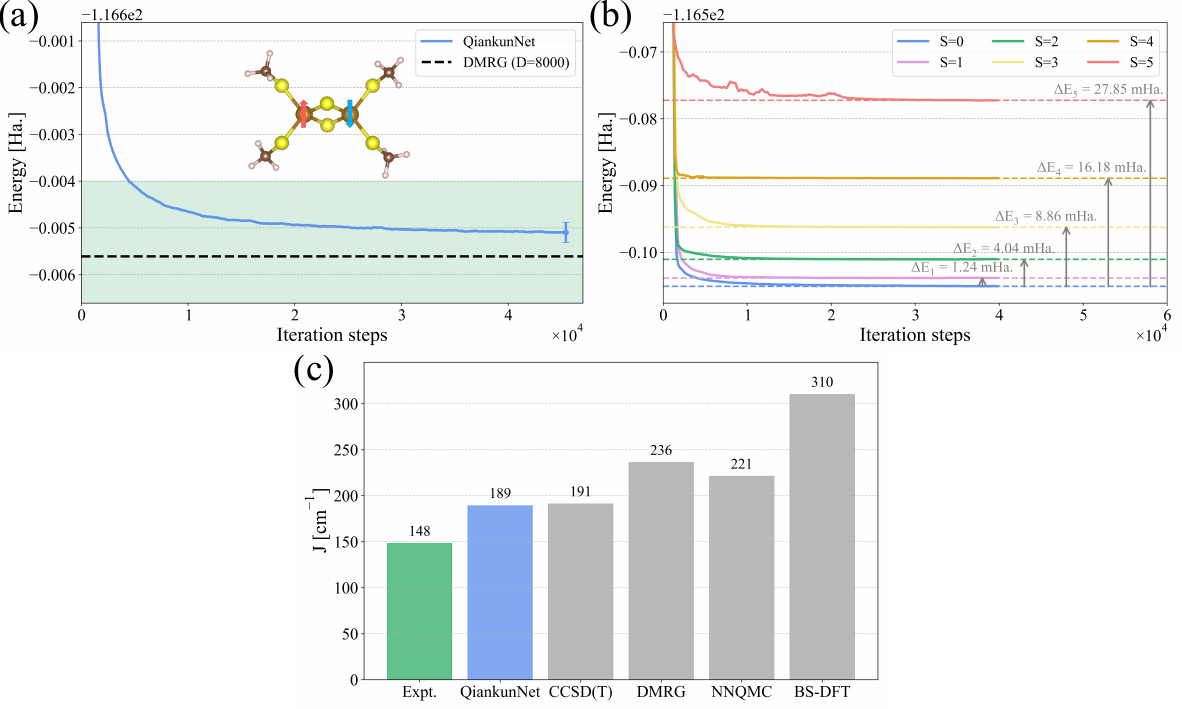}
    \caption{        
        Calculations for the \fesa\ complex $\left[\mathrm{Fe}_2 \mathrm{S}_2\left(\mathrm{SCH}_3\right)_4\right]^{2-}$ using a CAS(30e,20o) active space. The plotted energies represent the average values computed over the final 1000 steps of the training process. The green shaded area denotes the region within chemical accuracy (1 kcal/mol).
        (\textbf{a}) Optimization of the ground state energy using QiankunNet, compared with DMRG result. 
        (\textbf{b}) Optimization of different spin states, and their corresponding energy gap.
        (\textbf{c}) Magnetic exchange coupling constants $J$ derived from QiankunNet, in Comparison to Experimental Values and Other Computational Results.}
    \label{fig:fe2s2}
\end{figure}

\section*{Experimental results}

We have evaluated the performance of QiankunNet across a range of model and chemical systems, and found it to be particularly effective in treating strongly correlated systems, such as the Hubbard model and transition metal compounds. In this work, we demonstrate the accuracy and efficiency of QiankunNet using a classic and important family of chemical compound, the iron–sulfur clusters. These clusters are fundamental to biological electron transfer chains and numerous enzymatic catalysis pathways\cite{beinert1997iron}, with their functional versatility arising directly from complex electronic structures. As prototypical multi-iron centers linked by sulfide bridges, iron–sulfur clusters possess a high density of low-lying spin states\cite{venkateswara2004synthetic}. Near-degeneracy among the iron d-orbitals leads to significant electron correlation effects, rendering these systems markedly multi-configurational in nature\cite{sharma2014low}. The pronounced multi-reference character of these systems poses a fundamental challenge. Consequently, mean-field approaches like Hartree-Fock and DFT, along with many single-reference correlated methods, are generally inadequate for accurate description of iron-sulfur clusters.

To ensure a direct comparison with existing work, we adopt the same active space Hamiltonians as those used in the study by Li and Chan\cite{li2017spin}. These Hamiltonians, provided in FCIDUMP format, are defined on localized DFT orbitals for two iron-sulfur clusters: a \fesa\ cluster ($\left[\mathrm{Fe}_2 \mathrm{~S}_2\left(\mathrm{SCH}_3\right)_4\right]^{2-}$) and a \fesb\ cluster ($\left[\mathrm{Fe}_4 \mathrm{~S}_4\left(\mathrm{SCH}_3\right)_4\right]^{2-}$). For both systems, we utilize the published DMRG results for the S=0 state from the same reference as our benchmark.

\subsection*{\fesa\ Iron–sulfur cluster $\left[\mathrm{Fe}_2 \mathrm{~S}_2\left(\mathrm{SCH}_3\right)_4\right]^{2-}$}

For the \fesa\ cluster (30 electrons in 20 orbitals), DMRG with a very large bond
dimension (8000) yields a singlet ground-state energy of $-116.6056091$~Hartree.
We optimized a \system wavefunction for this $S=0$ state.
The optimization converged stably (Fig.~1a), and the final 1000 measurements were used to
estimate the energy. \system achieved a mean energy of
$-116.6051(2)$~Hartree, i.e.\ within $0.5$~millihartree of the DMRG value.
This small deviation (on the order of $1.3\times 10^{-3}$~eV) is well within
chemical accuracy, demonstrating the high fidelity of our ansatz for the ground state.

We next applied \system to various excited spin states of
\fesa\. By initializing and optimizing wavefunctions with total spin
$S=1,2,3,4,$ and $5$ (the maximal $S=5$ corresponding to high-spin alignment of
the two Fe(III) centers), we obtained an energy spectrum of low-lying spin states
(Fig.~1b). For each optimized state, we also computed the expectation values of
local and total spin operators. From the energies and $\langle \hat{S}^2 \rangle$
of these states, we can estimate the magnetic exchange coupling constant $J$
that characterizes the effective spin--spin interaction. We use the Yamaguchi~\cite{yamaguchi1986}
formula relating the energy difference between two spin states to their
$\langle \hat{S}^2 \rangle$ difference:
\begin{equation}
  J \;=\; \frac{E(S') - E(S)}{\langle \hat{S}^2 \rangle_{(S')} - \langle \hat{S}^2 \rangle_{(S)}}\,.
  \tag{1}
\end{equation}
Here we choose two representative states (e.g.\ the highest- and lowest-spin
states) for the estimation; in practice, we performed a linear fit using all six
spin states ($S=0$ through $5$) to extract a robust value of $J$ (Fig.~1c). \system  predicts an antiferromagnetic exchange coupling of
$J \approx 189~\mathrm{cm}^{-1}$ for \fesa\, with a strong linear correlation
($R^2 \approx 0.97$) indicating the validity of the spin-projection scheme.
Figure~1d compares this result to experimental estimates and previous computational
studies. Notably, our $J$ is in excellent agreement with the experimentally
inferred value and is closer to experiment than
the values obtained from high-level coupled-cluster\cite{schurkus2020theoretical}, DMRG\cite{sharma2014low}, or recently reported
NNQMC methods\cite{fu2025local}. In fact, among all methods compared, \system
produces the $J$ value closest to the experimental measurement, underscoring
the accuracy of our approach. We emphasize that our model achieved this accuracy
by optimizing each spin state's ground configuration separately---it did not
require explicit inclusion of excited-state configurations or a pre-defined
spin Hamiltonian. This suggests that the essential physics of exchange and
correlation in the \fesa\ cluster is well captured by the
\system ansatz within each spin sector.


\subsection*{\fesb\ Iron–sulfur cluster $\left[\mathrm{Fe}_4 \mathrm{~S}_4\left(\mathrm{SCH}_3\right)_4\right]^{2-}$}

We next consider the more complex \fesb\ cluster (54 electrons in 36 orbitals).
This system’s ground state is an \emph{antiferromagnetic} singlet resulting from coupling
two pairs of high-spin Fe centers. The reference DMRG energy for the $S=0$ state is
reported as $-327.2396$~Hartree. \system, with the CAS(54e,36o) active space, converged
to a mean energy of $-327.2398(7)$ Hartree for the singlet ground state. The \system{}
energy is statistically indistinguishable from the DMRG result (differing by only
$0.2$~mHartree, which is within the sampling uncertainty of $0.7$~mHartree). We can thus
conclude that \system{} achieves essentially the same ground-state energy accuracy as
DMRG for \fesb, again within chemical accuracy.

The spin correlations among the four Fe sites provide a more detailed fingerprint of the
\fesb ground state’s magnetic ordering. We computed the expectation values
$\langle \mathbf{S}_i \!\cdot\! \mathbf{S}_j \rangle$ for all distinct Fe--Fe pairs (Fig.~2b).
\system’s spin-correlation pattern is in quantitative agreement with earlier DMRG studies.
In particular, we observe that each Fe pairs antiferromagnetically with a specific partner
(yielding large negative correlations around $-3.9$ between certain pairs) while other pairs
have weaker ferromagnetic coupling (values around $+3.3$).
This pattern corresponds exactly to the lowest-energy spin configuration identified in DMRG,
where the four Fe atoms form two antiparallel pairs. Table~1 compares the numerical values
of these spin correlations from \system{} against the reference DMRG results. All values
agree within a small margin, validating that \system{} captures the detailed magnetic
structure of the \fesb cluster. We note that such spin properties are higher-order
observables not directly targeted during energy optimization, yet our wavefunction reproduces
them accurately---a testament to its quality.

\begin{figure}
    \centering
    \includegraphics[width=1.0\linewidth]{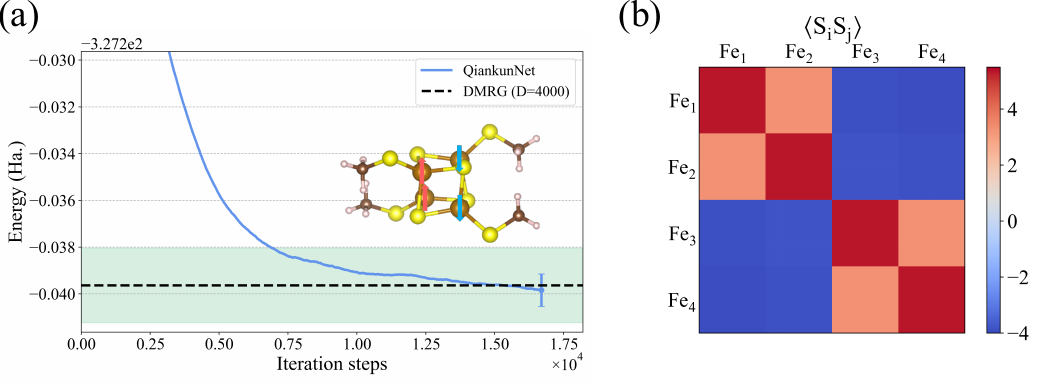}
    \caption{Calculations for the \fesb\ complex $\left[\mathrm{Fe}_4 \mathrm{~S}_4\left(\mathrm{SCH}_3\right)_4\right]^{2-}$. 
    (\textbf{a}) QiankunNet optimization for ground state of \fesb\ with an active space of CAS(54e,36o). Ensemble averages of the energies were obtained from the last 1000 steps of the training trajectory. Chemical accuracy (1 kcal/mol) is demarcated by the green shaded region.
    (\textbf{b}) Calculated spin correlation between 4 iron atoms.
    }
    \label{fig:fe4s4}
\end{figure}

\begin{table}[th]
    \centering
    \renewcommand{\arraystretch}{1.5}
    \begin{tabular}{ccc}
    \hline
    \hline
    $\langle \hat{S}_i\cdot  \hat{S}_j \rangle$  & \system   &  DMRG\cite{sharma2014low} \\
    \hline
    $\mathrm{Fe_1-Fe_1}$    &  5.33 &  5.27 \\
    $\mathrm{Fe_2-Fe_2}$    &  5.35 &  5.26 \\
    $\mathrm{Fe_3-Fe_3}$    &  5.33 &  5.27 \\
    $\mathrm{Fe_4-Fe_4}$    &  5.35 &  5.27 \\
    $\mathrm{Fe_1-Fe_2}$    &  3.31 &  3.24 \\
    $\mathrm{Fe_1-Fe_3}$    & -3.89 & -4.05 \\
    $\mathrm{Fe_1-Fe_4}$    & -3.96 & -4.05 \\
    $\mathrm{Fe_2-Fe_3}$    & -3.84 & -4.05 \\
    $\mathrm{Fe_2-Fe_4}$    & -3.88 & -4.04 \\
    $\mathrm{Fe_3-Fe_4}$    &  3.31 &  3.24 \\
    \hline
    \end{tabular}
    \caption{Spin-spin correlations ($\langle \hat{S}_i\cdot  \hat{S}_j \rangle$) between iron atoms in the \fesb\ cluster, compared with DMRG results from Ref. \cite{sharma2014low}.}
    \label{tab:fe4s4_spipn}
\end{table}

\section*{Discussion} 

In this study, we introduce a scheme, which integrates Transformer architectures with the neural network backflow framework to address the challenge of strongly correlated quantum systems. This represents a significant advance in adapting language-model-inspired architectures to capture the complex correlation patterns of electrons — the “grammar” of many-body quantum mechanics. Just as Transformers have redefined natural language processing and accelerated breakthroughs in protein modeling, QiankunNet demonstrates their capacity to reshape our understanding of electronic correlations in transition-metal clusters and beyond.

We acknowledge the seminal contributions of neural network backflow methods~\cite{Luo2019,Liu2024-1,liu2025NNBF}, which established the power of configuration-dependent orbital transformations for improving variational wavefunctions. QiankunNet extends this framework by embedding backflow directly within a Transformer backbone, thereby combining self-attention’s ability to capture global orbital correlations with the systematic orbital reparameterization of backflow. This synergy enables us to tackle challenging systems such as \fesa\ and \fesb\, where conventional methods struggle due to severe static correlation and dense spin manifolds.

More broadly, our work establishes a conceptual bridge between modern deep learning architectures and physics-inspired wavefunction ansätze. By combining parallelizable Transformer layers, and determinant-based amplitudes, QiankunNet achieves a rare blend of scalability and accuracy. QiankunNet targets realistic molecular active spaces and explicitly encodes many-body correlations via backflow, rather than relying purely on large data-driven training. The result is a framework that balances expressive power with physical interpretability, potentially allowing insight into the correlated orbital patterns it learns.

An exciting implication of this work is the prospect of transferable, pretrainable models for quantum chemistry. One can envision training a Transformer model on a diverse set of molecules or materials to serve as a “foundation” wavefunction that can be fine-tuned to new systems, much as language models are fine-tuned for specific tasks. QiankunNet moves toward this vision by demonstrating that a single neural architecture can smoothly handle both weakly and strongly correlated regimes – from simple reference states to highly multi-reference clusters – within one unified ansatz. In future work, incorporating techniques like transfer learning or unsupervised pre-training on generated data may further enhance the model’s generality.

In conclusion, QiankunNet provides a new tool for strongly correlated electronic structure, combining the latest advances in deep learning with respected quantum chemistry techniques. Its ability to achieve high accuracy in challenging cases (like iron–sulfur clusters) using tractable computational resources is a promising step toward reliable and scalable many-electron calculations. We anticipate that QiankunNet will become an important component in the toolbox of computational chemists and physicists, particularly for materials and bioinorganic complexes where traditional wavefunction methods face intrinsic limitations. By continuing to refine such neural-network ansätze and exploring their limits, we inch closer to the long-term goal of unified, foundation-model approaches that can seamlessly tackle the diverse correlation problems encountered across chemistry and materials science.

\section*{Methods}
\subsection*{The framework}

QiankunNet leverages the strong expressive power of the transformer architecture to construct configuration-dependent orbitals—referred to as backflowed single-particle orbitals (SPOs)—whose matrix determinant yields the configuration coefficient. In general, the wavefunction ansatz of QiankunNet takes the form:
\begin{equation}
\label{QK-BF}
\psi(\boldx)=\det\big(\widetilde{A}_{\theta}(\boldx)[\boldx]\big).
\end{equation}
The neural network responsible for generating the configuration-dependent orbital matrix $\widetilde{A}_{\theta}(\boldx)$ comprises three primary components: embedding of occupation strings, feature extraction, and orbital mapping. We now elaborate on each of these components in turn.

The input to QiankunNet is a many-body configuration represented by occupation strings: $|\boldx\rangle = |x_1, x_2, \ldots, x_N\rangle$, where each $x_i \in {0,1}$ indicates the occupancy of the $i$-th spin orbital. We begin by partitioning the configuration into groups of spin-orbitals, with each group comprising $t$ consecutive spin-orbitals treated as a single token. This grouping results in $N_t = \lceil N_{so} / t \rceil$ tokens, where $N_{so}$ denotes the total number of spin-orbitals. If $N_{so}$ is not divisible by $t$, the configuration is padded with zeros at the end to form complete tokens (note that this padding is applied only for embedding purposes).

Each token represents a local occupancy pattern over $t$ spin-orbitals, corresponding to one of $2^t$ possible occupancy states within that subspace. Each such state is mapped to a $d_f$-dimensional feature vector. For the $i$-th token, the embedding is performed via a token-specific matrix $E_i \in \mathbb{R}^{2^t \times d_f}$, whereby the actual occupancy pattern of the token selects a specific row $X_i \in \mathbb{R}^{d_f}$ from $E_i$. In QiankunNet, each token employs a dedicated embedding matrix, resulting in a comprehensive embedding structure $E \in \mathbb{R}^{N_t \times 2^t \times d_f}$. After embedding, the configuration $|\boldx\rangle$ is transformed into a feature matrix $X \in \mathbb{R}^{N_t \times d_f}$.

Following the embedding process, a given configuration is mapped into a feature matrix $X$, where each token is embedded independently. This matrix $X$ then serves as the initial input, denoted as $X^{(0)}$, to a stack of Transformer encoder layers designed to capture complex dependencies among tokens.

Each encoder layer $l \in {1, 2, \dots, L_e}$ follows the standard Transformer architecture and comprises a multi-head self-attention (MHA) mechanism followed by a feed-forward network (FFN), with residual connections and layer normalization applied after each sub-layer. The computation within each layer proceeds as follows:

\begin{equation}
\begin{aligned}
H^{(l-1)} &= \text{LayerNorm}\left( X^{(l-1)} + \text{MHA}^{(l)}\left(X^{(l-1)}\right) \right), \\
X^{(l)} &= \text{LayerNorm}\left( H^{(l-1)} + \text{FFN}^{(l)}\left(H^{(l-1)}\right) \right).
\end{aligned}
\end{equation}

Here, the multi-head attention module $\text{MHA}^{(l)}$ operates on the input representations $X^{(l-1)}$ by projecting them into queries ($Q$), keys ($K$), and values ($V$) via learned weight matrices:

\begin{equation}
    Q=X^{(l-1)} W_Q^{(l)}, \quad K=X^{(l-1)} W_K^{(l)}, \quad V=X^{(l-1)} W_V^{(l)}
\end{equation}

The attention output is computed as a weighted sum over heads:
\begin{equation}
\text{MHA}^{(l)}(Q, K, V) = \sum_{h=1}^{N_h} \text{softmax}\left( \frac{Q_h K_h^\top}{\sqrt{d_{\text{atten}} / N_h}} \right) V_h,
\end{equation}
where $Q_h$, $K_h$, and $V_h$ are the partitioned sub-matrices corresponding to the $h$-th attention head, $N_h$ is the total number of attention heads, and $d_{\text{atten}}$ is the hidden dimension within the attention mechanism.

Subsequent to the multi-head attention module, each encoder layer incorporates a feed-forward network (FFN) that applies further non-linear processing to the outputs of the attention mechanism. The FFN consists of a linear transformation followed by a ReLU activation function, expressed as:

\begin{equation}
    \text{FFN}^{(l)}\left( H^{(l-1)}\right) = \operatorname{ReLU} \left( H^{(l-1)}W^{(l)}_{\text{FFN}} +b^{(l)}_\text{FFN}\right)
\end{equation}

This sub-layer enhances the representational capacity of the model by independently transforming each token representation through the same parameterized function.

After processing through the $L_e$ encoder layers, the dependencies among different tokens are effectively captured, yielding an output feature matrix $\tilde{X} \in \mathbb{R}^{N_t \times d_f}$:

\begin{equation}
\tilde{X} =
\begin{bmatrix}
\tilde{X}_1 \\
\tilde{X}_2 \\
\vdots \\
\tilde{X}_{N_t}
\end{bmatrix},
\end{equation}

where each token is represented by a feature vector $\tilde{X}_i \in \mathbb{R}^{d_f}$. The final component of the QiankunNet architecture is the orbital mapping module, which transforms the feature vector of each token into the corresponding rows of the single-particle orbital (SPO) matrix $\widetilde{A}_\theta(\mathbf{x})$. This is achieved via a dedicated multi-layer perceptron (MLP) block for each token:

\begin{equation}
\widetilde{A}_\theta(\mathbf{x}) =
\begin{bmatrix}
\widetilde{A}_{\theta,1}(\mathbf{x})\\
\widetilde{A}_{\theta,2}(\mathbf{x})\\
\vdots\\
\widetilde{A}_{\theta,N_t}(\mathbf{x})
\end{bmatrix}
=
\begin{bmatrix}
\operatorname{MLP}_1 \left( \tilde{X}_1 \right)\\
\operatorname{MLP}_2 \left( \tilde{X}_2 \right)\\
\vdots\\
\operatorname{MLP}_{N_t} \left( \tilde{X}_{N_t} \right)
\end{bmatrix}.
\end{equation}

Here, each $\widetilde{A}_{\theta,i}(\mathbf{x}) \in \mathbb{R}^{t \times N_e}$ is a sub-matrix of the full SPO matrix $\widetilde{A}_{\theta}(\mathbf{x}) \in \mathbb{R}^{N_{so} \times N_e}$, containing the rows corresponding to the spin-orbitals within the $i$-th token. The MLP blocks share the same architecture across tokens but are parameterized independently. Each MLP consists of $L_m$ linear layers, with each layer $l^\prime \in {1, 2, \dots, L_m}$ employing a ReLU activation function:

\begin{equation}
\tilde{X}^{(l^{\prime})}_n = \operatorname{ReLU} \left( \tilde{X}^{(l^{\prime}-1)}_n W^{(l^\prime)}_n + b^{(l^\prime)}_n \right).
\end{equation}

A final linear output layer is appended to project the transformed features into the SPO matrix:

\begin{equation}
\widetilde{A}_{\theta,n}(\mathbf{x}) = \tilde{X}^{(L_m)}_n W_{o,n} + b_{o,n},
\end{equation}

where $W_{o,n} \in \mathbb{R}^{d_{\text{mlp}} \times (t \times N_e)}$ and $b_{o,n} \in \mathbb{R}^{t \times N_e}$ are the output layer weights and biases, respectively. This operation maps the features of the $n$-th token to $t$ contiguous rows of $\widetilde{A}_{\theta}(\mathbf{x})$. The configuration coefficient is then obtained by selecting the occupied rows of $\widetilde{A}_{\theta}(\mathbf{x})$ to form a $N_e \times N_e$ matrix $\widetilde{A}_{\theta}(\mathbf{x})[\mathbf{x}]$, whose determinant is evaluated as described in Eq. \ref{QK-BF}.

The above describes the construction of a single SPO. For strongly correlated systems, however, a single determinant is often insufficient to achieve chemical accuracy, necessitating a more expressive wavefunction ansatz. Following the progression from Slater–Jastrow to multi-Slater–Jastrow wavefunctions—and in line with previous neural network backflow (NNBF) studies\cite{Liu2024-1,liu2025NNBF}—we enhance the expressiveness of QiankunNet by employing multiple SPOs. The resulting wavefunction is given by:

\begin{equation}
\label{QK_BF_multi_determinant}
\psi(\mathbf{x}) = \sum_{d=1}^{D} \det\left( \widetilde{A}^{d}_{\theta}(\mathbf{x})[\mathbf{x}] \right),
\end{equation}

where $\widetilde{A}^{d}_{\theta}(\mathbf{x})$ denotes the $d$-th SPO matrix generated by the model and $D$ is the total number of determinants. To produce $D$ distinct SPOs, the output layer of each MLP block is extended such that $W_{o,n} \in \mathbb{R}^{d_{\text{mlp}} \times (D \cdot t \cdot N_e)}$ and $b_{o,n} \in \mathbb{R}^{D \cdot t \cdot N_e}$. The output of each MLP is then reshaped into $D$ separate SPO sub-matrices.

\subsection*{General QiankunNet setting used in this study}
Tab. \ref{tab:hyperparameter_qkbf} summarizes all hyperparameters associated with the neural network, including their descriptions and default values.

\begin{table}[htb]
    \centering
    \begin{tabular}{c c c}
    \hline
    \hline
    Notation &  Description&  Default Value\\
    \hline
    $N_o$    &  Number of orbitals &  system dependent \\
    $N_{so}$   &  Number of spin-orbitals &   $2 \times N_o$\\
    $N_e$    &  Number of electrons &  system dependent \\
    $t$     &  spin-orbitals in one token &  4 \\
    $N_t$     &  Number tokens &  $\lceil N_{so} / t \rceil$ \\
    $d_f$      &  Dimension of feature vector  &   256 \\
    $L_{e}$  &  Transformer encoder layers  &   2 \\
    $N_{h}$  &   Attention heads  &   4 \\
    $d_{atten}$  &   hidden feature of attention layer  &   256 \\
    $L_{m}$  &  MLP layers for orbital mapping  &   2 \\
    $d_{mlp}$  &  Hidden features in MLP blocks  &   256 \\
    $D$  &  Number of SPOs to be constructed  &   2 \\
    \hline
    \end{tabular}
    \caption{Summary of the hyperparameters, descriptions, and default values used in the QiankunNet architecture.}
    \label{tab:hyperparameter_qkbf}
\end{table}

\subsection*{Variational Monte Carlo}

The QiankunNet wavefunction is optimized with Variational Monte Carlo (VMC) method. The expectation value within the VMC framework is calculated with a stochastic sampling scheme instead of calculated variationally:
\begin{equation}
\label{eq:local_estimator}
    \langle \hat{O}\rangle
    = \frac{\langle \Psi | \hat{O} | \Psi \rangle}
       {\langle \Psi | \Psi \rangle}
    \approx \mathbb{E}_{\boldx \sim p(\boldx)}\left[O_{\text{loc}}(\boldx)\right]
    \equiv \bar{O}_{\text{loc}}
\end{equation}
$p(\boldx)$ is the probability distribution from which configuration $\boldx$ is sampled, and obeys:
\begin{equation}
\label{eq:prob}
    p(\boldx) = \frac{\psi^2(\boldx)}{\sum_{\boldx^{\prime}} \psi^2(\boldx^{\prime})}
\end{equation}
with $\psi(\boldx) = \langle \boldx|\Psi\rangle$. $O_{\text{loc}}(\boldx)$ is known as the local estimator:
\begin{equation}
    O_{\text{loc}}(\boldx) = \frac{\langle \boldx | \hat{O}|\Psi \rangle}{\left\langle \boldx | \Psi \right\rangle} = \sum_{\boldx^{\prime}}O_{\boldx\boldx^{\prime}}\frac{\psi(\boldx^{\prime})}{\psi(\boldx)}
\end{equation}
$O_{\boldx\boldx^{\prime}} = \langle \boldx|\hat{O}|\boldx^{\prime}\rangle$ is the matrix element of $\hat{O}$. The VMC optimization of QiankunNet depends on the estimation of energy expectation:
\begin{equation}
\label{eq:local_energy}
\begin{aligned}
    \bar{E}_\theta = &\mathbb{E}_{\boldx \sim p(\boldx)}\left[E_{loc}(\boldx)\right] \\
    E_{\text{loc}}(\boldx) &= \sum_{\boldx^{\prime}}H_{\boldx\boldx^{\prime}}\frac{\psi_\theta(\boldx^{\prime})}{\psi_\theta(\boldx)}
\end{aligned}
\end{equation}

Here, $E_{\text{loc}}$ is termed the local energy. The optimization procedure updates the parameters $\theta$ according to
\begin{equation}
    \theta_{new} \leftarrow \theta - \eta \cdot \mathbf{g}
\end{equation}
where $\eta$ is the learning rate and $\mathbf{g}$ denotes the gradient. The specific form of $g$ depends on the choice of optimizer. 

As detailed in the preceding section, the optimization of QiankunNet is performed using gradient-based methods. This section describes the specific optimizers employed and outlines the procedure for obtaining the requisite gradients. The overall training process utilizes two distinct optimizers: the widely adopted Adam optimizer \cite{kingma2014adam} and the recently proposed MARCH optimizer by Gu et al. \cite{gu2025solving}.

Within the stochastic gradient descent (SGD) framework, the gradient for VMC with respect to the parameters $\theta$, based on sampled local energies, is given by:

\begin{equation}
\label{eq:gradient_sgd}
\mathbf{g} = 2 \cdot \mathbb{E}_{\mathbf{x}\sim p_{\theta}(\mathbf{x})}\Bigl[
\bigl( E_{\text{loc}}(\mathbf{x}) - \mathbb{E}_{\mathbf{x}\sim p_{\theta}(\mathbf{x})}[E_{\text{loc}}(\mathbf{x})] \bigr)
\mathbf{O}(\mathbf{x})
\Bigr]
\end{equation}

Here, $\mathbf{O}$ denotes the Jacobian matrix of the logarithmic wavefunction, $\ln \psi_{\theta}$, with dimensions $\mathbb{R}^{B \times N}$, where $B$ is the batch size (number of sampled configurations $\mathbf{x}$) and $N$ is the number of parameters in the QiankunNet network. The matrix elements of $\mathbf{O}$ are defined as:

\begin{equation}
\label{eq:gradient_O}
O_{\theta}(\mathbf{x}) = \frac{\partial \ln\psi_{\theta}(\mathbf{x})}{\partial\theta}
\end{equation}

The Adam optimizer utilizes this gradient estimate (Eq. \ref{eq:gradient_sgd}) in conjunction with two momentum-based estimators to adaptively adjust the learning rate for each parameter, facilitating more efficient convergence. \cite{kingma2014adam} 

The optimization of the QiankunNet wavefunction employs a two-stage strategy. In the initial stage, the Adam optimizer is utilized due to its computational efficiency and low memory footprint. However, while Adam facilitates rapid initial convergence, it typically fails to achieve a wavefunction that meets chemical accuracy requirements. Therefore, once the Adam optimizer has reached a preliminary optimization plateau, the MARCH optimizer \cite{gu2025solving}—a second-order method—is employed to refine the wavefunction to the desired precision.
The MARCH optimizer \cite{gu2025solving} belongs to the family of stochastic reconfiguration (SR) methods \cite{SR0,SR1,SR2,SR3} and is derived from the MinSR optimizer \cite{MinSR,MinSR_parallel}. SR-based approaches optimize the neural network quantum state within the variational manifold, which is mathematically equivalent to imaginary time evolution under the time-dependent variational principle \cite{SR0}.

The MinSR gradient is given by:
\begin{equation}
    \mathbf{g} = \mathbf{O}^{T} \left( \mathbf{O}\mathbf{O}^{T} + \lambda \mathbf{I}\right)^{-1} \chi
\end{equation}
where $\mathbf{S}=\mathbf{O}\mathbf{O}^{T}$ is referred to as the Fisher matrix and $\lambda$ serves as a regularization parameter to prevent matrix singularity.$\chi$ denotes the precomputed local energy term defined as
\begin{equation}
    \chi = 2\left(E_{\text{loc}}(\mathbf{x}) - \mathbb{E}_{\mathbf{x}\sim p_{\theta}(\mathbf{x})}[E_{\text{loc}}(\mathbf{x})]\right)
\end{equation}

The MARCH optimizer extends the MinSR approach by incorporating two momentum terms to guide the optimization trajectory, enabling adaptive, per-parameter learning rate adjustments.

The first momentum term, denoted as $\mathbf{m}_t \in \mathbb{R}^{N}$, accumulates historical gradient information with an exponential decay factor: $\mathbf{m}_t = \beta_1\mathbf{g}_{t-1}$. The second momentum term, $\mathbf{v}_t \in \mathbb{R}^{N}$, tracks the velocity of gradient changes according to $\mathbf{v}_t = \beta_2\mathbf{v}_{t-1} + (\mathbf{g}_{t-1} - \mathbf{g}_{t-2})^{2}$, thereby capturing the relative optimization rates across different parameters. This velocity estimate is subsequently utilized to scale the learning rate individually for each parameter.

The resulting gradient update for the MARCH optimizer is given by:
\begin{equation}
\mathbf{g}_t = \operatorname{diag}\left(\mathbf{v}_{t}\right)^{-1/2} \mathbf{O}^T\left(\mathbf{O} \operatorname{diag}\left(\mathbf{v}_{t}\right)^{-1/2} \mathbf{O}^T + \lambda \mathbf{I}\right)^{-1}\left(\chi - \mathbf{O} \mathbf{m}_{t}\right) + \mathbf{m}_{t}
\end{equation}

To enhance the stability of the MARCH optimizer, two stabilization techniques are employed: gradient clipping and a norm constraint on the update step.

A clipping operation is applied to the momentum term $\mathbf{v}_k$ to mitigate numerical instability. In cases where certain parameters exhibit consistently zero gradients, the corresponding components of $\mathbf{v}_{t,\theta}$ can become exceedingly small, causing $\mathbf{v}_t^{-1/2}$ to produce numerical overflow and disrupt the optimization. The clipping prevents these values from falling below a predefined threshold.
\begin{equation}
    \mathbf{v}^{\text{clip}}_t =\text{min}(\text{max}(\mathbf{v}_t,\frac{1}{\epsilon}),\epsilon)
\end{equation}

Furthermore, a norm constraint is imposed on the gradient $\mathbf{g}_k$ at each optimization step. The effective learning rate for the update is determined by:

\begin{equation}
\eta_{\text{eff}} = \min\left(\eta, \frac{c}{||\mathbf{g}_t||_2}\right)
\end{equation}

This ensures that the magnitude of the parameter update remains bounded, thereby improving training stability. Tab.\ref{tab:hyperparameter_MARCH} summarizes all the hyper-parameters used in MARCH optimizer.

\begin{table}[htb]
    \centering
    \begin{tabular}{c c c}
    \hline
    \hline
    Notation &  Description&  Default Value\\
    \hline
    $N$    &  Number of parameters &  system dependent \\
    $B$    &  Batch size &  4096 \\
    $\eta$    &  Learning rate &  0.1 \\
    $\beta_1$ &  Exponential decay rate 1 &  0.95 \\
    $\beta_2$ &  Exponential decay rate 2 &  0.995 \\
    $\lambda$ &  Regularization term &  0.001 \\
    $\epsilon$ &  Clipping threshold for $\mathbf{v}_t$ &  $1\times 10^8$ \\
    $c$       &  Norm constraint &  0.1 \\
    \hline
    \end{tabular}
    \caption{Summary of the hyperparameters used in MARCH optimizer}
    \label{tab:hyperparameter_MARCH}
\end{table}



Stochastic reconfiguration (SR)-based optimizers are generally characterized by high memory consumption and computational cost, primarily due to the construction and manipulation of the $\mathbf{O}$ matrix. To mitigate these constraints and accelerate the MARCH optimizer, a data parallelization scheme is implemented to distribute the computational workload across multiple GPUs. This approach follows the distributed framework for the MinSR optimizer developed by Rende et al. \cite{MinSR_parallel}.

\begin{figure}[th]
\centering
\includegraphics[width=0.95\linewidth]{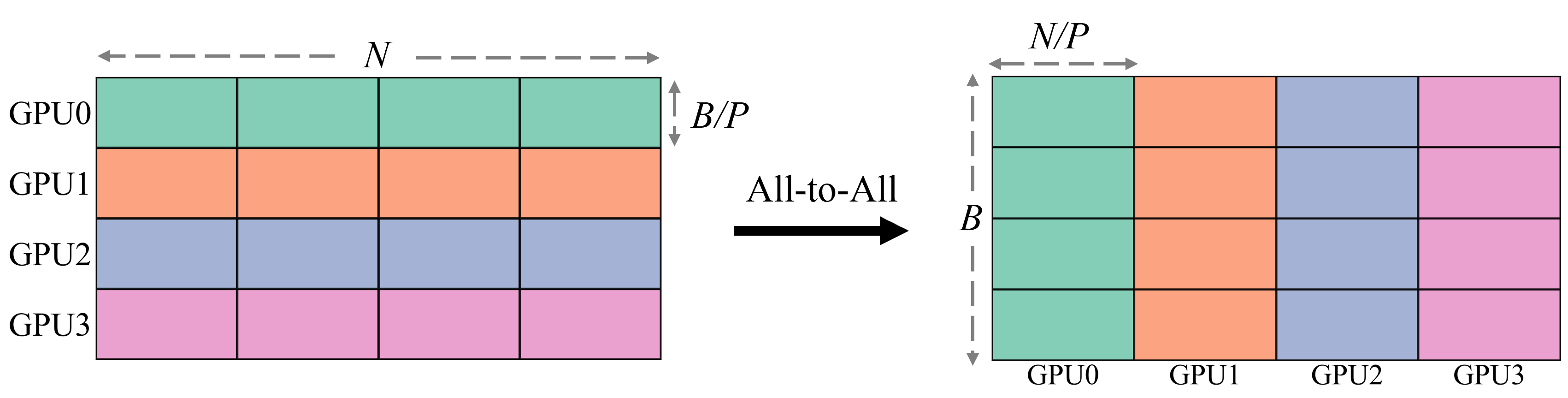}
\caption{Schematic of the all-to-all collective communication used to transform the $\mathbf{O}$ matrix from a row-wise parallel to a column-wise parallel distribution across GPUs.}
\label{fig:all_to_all}
\end{figure}

The parallelization strategy involves distributing the rows of the Jacobian matrix $\mathbf{O}$ across available GPUs. To facilitate subsequent matrix operations that require column-wise access, an all-to-all collective communication step is performed. This operation redistributes the matrix elements such that each GPU stores a distinct set of columns, transforming the data layout from row-parallel to column-parallel, as illustrated in Fig. \ref{fig:all_to_all}.

The row-wise parallelized segment of the matrix $\mathbf{O}$ assigned to each GPU is denoted as $\mathbf{O}_{\text{row}}^{\text{local}}$. Following the all-to-all communication, the resulting column-wise distributed segment is correspondingly denoted as $\mathbf{O}_{\text{col}}^{\text{local}}$.
This distribution scheme allows the large matrix $\mathbf{O} \in \mathbb{R}^{B\times N}$to be efficiently partitioned across $P$ GPUs, with each GPU storing either $\mathbf{O}_{\text{row}}^{\text{local}}\in\mathbb{R}^{B/P \times N}$ or $\mathbf{O}_{\text{col}}^{\text{local}} \in \mathbb{R}^{B \times N/P}$. Other key components of the MARCH optimizer are similarly distributed: the gradient $\mathbf{g}^{\text{global}}_t \in \mathbb{R}^{N}$ and momentum terms $\mathbf{m}^{\text{global}}_t ,\mathbf{v}^{\text{global}}_t \in \mathbb{R}^{N}$ are partitioned such that each GPU maintains information for only $N/P$ parameters, i.e. $\mathbf{g}_t \in \mathbb{R}^{N/P}$, $\mathbf{m}_t,\mathbf{v}_t \in \mathbb{R}^{N/P}$. All matrix operations are performed in a distributed manner, coordinated through collective communication primitives. The complete implementation of the distributed MARCH optimizer is outlined in the pseudocode provided in Algorithm 1.

\begin{algorithm}
\label{algo:march}
\caption{Distributed MARCH optimizer}
\renewcommand{\algorithmicrequire}{\textbf{Input:}}
\renewcommand{\algorithmicensure}{\textbf{Output:}}
    \begin{algorithmic}[1]
    \REQUIRE Neural network parameters $\theta$; number of parameters $N$; batch size $B$; number of processes (world size) $P$; local rank $r$; learning rate $\eta$; decay rate $\beta_1,\beta_2$; regularization term $\lambda$; clipping threshold $\epsilon$; norm constraint $c$.
    \ENSURE optimized neural network parameters $\theta$
    \STATE \textbf{Initialize}: $\mathbf{g}_0 \gets \mathbf{0}_{N/P}$, $\mathbf{v}_0 \gets \mathbf{1}_{N/P}$      
    \FOR{$t=1$ to MaxIter}
        \STATE Sample $B/P$ configurations $\mathbf{x}$
        \STATE $\ln \psi_\theta \gets \text{QiankunNet}_{\theta}(\mathbf{x})$
        \STATE $ \chi^{\text{local}} \gets 2\left(E_{\text{loc}}(\mathbf{x}) - \text{AllReduce}(\mathbb{E}[E_{\text{loc}}(\mathbf{x})], \text{AVG} )\right)$    \quad \text{   // Local energy calculation}
        \STATE $\chi \gets \text{AllGather}(\chi^{\text{local}})$
        \STATE $\mathbf{O}_{\text{row}}^{\text{local}} \gets \nabla_\theta \ln\psi_{\theta}$  \quad \quad \text{// Compute gradients via automatic differentiation}
        \STATE $\mathbf{O}_{\text{row}}^{\text{local}} \gets \mathbf{O}_{\text{row}}^{\text{local}} - \text{AllReduce}(\mathbb{E}[\mathbf{O}_{\text{row}}^{\text{local}}], \text{AVG}) $  \quad \text{// Gradient centering}
        \STATE $\mathbf{O}^{\text{local}}_{\text{col}} \xleftarrow{\text{AlltoAll}} \mathbf{O}^{\text{local}}_{\text{row}}$
        \STATE $\mathbf{S}^{\text{local}} \gets \mathbf{O}^{\text{local}}_{\text{col}} \cdot \text{diag}(\mathbf{v}^{-1/2}) \cdot (\mathbf{O}^{\text{local}}_{\text{col}})^T$   \quad \text{// Compute local Fisher matrix}
        \STATE $\mathbf{S} \gets \text{AllReduce}(\mathbf{S}^{\text{local}}, \text{SUM}) + \lambda \mathbf{I}$  \quad \quad \text{// Construct global Fisher matrix}
        \STATE $\xi \gets \chi - \beta_1 \cdot \text{Allreduce}(\mathbf{O}^{\text{local}}_{\text{col}} \cdot \mathbf{g}_{t-1},\text{SUM})$
        \STATE $\zeta \gets \text{argmin}_x \|\mathbf{S}x - \xi\|^2$ \quad \text{// Solve linear equation $\mathbf{S}x = \xi$ with least-square method}
        \STATE $\mathbf{g}_{t} \gets \text{diag}( \mathbf{v}^{-1/2}) (\mathbf{O}^{\text{local}}_{\text{col}})^T \zeta  + \beta_1 \mathbf{g}_{t-1}$   \quad \text{// Compute local gradient}
        \STATE $\mathbf{v}_{t+1} \gets \beta_2 \mathbf{v}_t + (\mathbf{g}_t - \mathbf{g}_{t-1})^2$  \quad \text{// Compute next step $\mathbf{v}$}
        \STATE $\mathbf{v}_{t+1} \gets \text{min}(\text{max}(\mathbf{v}_{t+1},\frac{1}{\epsilon}),\epsilon)$ \quad \text{// Clip $\mathbf{v}$}
        \STATE  $\mathbf{g}^{\text{global}}_t \gets \text{AllGather}(\mathbf{g}_t) $  \quad \text{// Gather local gradient to global gradient}
        \STATE $\theta \gets \theta - \min\left(\eta, \frac{c}{||\mathbf{g}^{\text{global}}_t||_2}\right) \cdot \mathbf{g}^{\text{global}}_t$ \quad \text{// Update $\theta$ with global gradient}
    \ENDFOR
    \RETURN $\theta$
    \end{algorithmic}
\end{algorithm}

An essential component of the VMC method is the efficient sampling of configurations $\mathbf{x}$, from the probability distribution $p_\theta(\mathbf{x})$ defined in Eq. \ref{eq:prob}. In this work, configurations are generated using the Markov Chain Monte Carlo (MCMC) method, specifically employing the Metropolis-Hastings algorithm.

The sampling is constrained to a specific subspace of the full Hilbert space, defined by fixed electron number $N_e$ and total spin projection $S_z$. This fixes the number of spin-up ($N_{\uparrow} = N_e/2 + S_z$) and spin-down ($N_{\downarrow} = N_e/2 - S_z$) electrons. This restriction is explicitly enforced throughout the MCMC sampling procedure.

To implement the Metropolis algorithm, an ensemble of Markov chains is initialized with random configurations satisfying the above constraints. The chains are then evolved iteratively. At each step, for a chain in the current configuration $\mathbf{x}_i$, a new candidate configuration $\mathbf{x}_j$ is proposed. The acceptance of this proposal is determined by the Metropolis acceptance criterion:
\begin{equation}
P(\mathbf{x}_i \to \mathbf{x}_j) = \min \left[ 1, \frac{|\psi_\theta(\mathbf{x}_j)|^2}{|\psi_\theta(\mathbf{x}_i)|^2} \right].
\end{equation}
A uniformly distributed random number $r \in [0, 1)$ is generated; if $r < P(\mathbf{x}_i \to \mathbf{x}_j)$, the proposed move is accepted and the chain's state is updated to $\mathbf{x}_j$. Otherwise, the move is rejected, and the chain remains in state $\mathbf{x}_i$.

In this study, a "random-hopping" strategy is adopted for proposing new configurations. This scheme involves selecting, at random, an electron from an occupied orbital and moving it to a randomly chosen unoccupied orbital of the same spin. This single-electron excitation constitutes one Monte Carlo proposal, providing a simple and computationally efficient method for exploring the constrained configuration space.



\subsection*{Spin Measurement}
In the VMC framework, the expectation value of a spin operator is computed by first constructing the target operator $\hat{\mathbf{S}}$ and then evaluating its expectation value using the local estimator method presented in Eq. \ref{eq:local_estimator}.

The basic spin operators can be written:
\begin{equation}
\begin{aligned}
\hat{S}_i^x & =\frac{1}{2}\left(\hat{c}_{i \uparrow}^{\dagger} \hat{c}_{i \downarrow}+\hat{c}_{i \downarrow}^{\dagger} \hat{c}_{i \uparrow}\right) \\
\hat{S}_i^y & =\frac{i}{2}\left(\hat{c}_{i \downarrow}^{\dagger} \hat{c}_{i \uparrow}-\hat{c}_{i \uparrow}^{\dagger} \hat{c}_{i \downarrow}\right) \\
\hat{S}_i^z & =\frac{1}{2}\left(\hat{c}_{i \uparrow}^{\dagger} \hat{c}_{i \uparrow}-\hat{c}_{i \downarrow}^{\dagger} \hat{c}_{i \downarrow}\right)\\
\hat{S}_i^{+} & = \hat{S}^x_i +i\hat{S}^y_i=\hat{c}_{i \uparrow}^{\dagger} \hat{c}_{i \downarrow} \\
\hat{S}_i^{-} & = \hat{S}^x_i -i\hat{S}^y_i=\hat{c}_{i \downarrow}^{\dagger} \hat{c}_{i \uparrow} \\
\hat{S}_i \hat{S}_j  &=\hat{S}_i^x \hat{S}_j^x+\hat{S}_i^y \hat{S}_j^y+\hat{S}_i^z \hat{S}_j^z
\end{aligned}
\end{equation}
To measure the expectation value of $\hat{S}^2$ of a target orbital or orbital set $\mathcal{P}$:
\begin{equation}
    \hat{S}^2_{\mathcal{P}} = \sum_{i \in \mathcal{P}}\sum_{j\in \mathcal{P},j > i} 2\hat{S}_i\hat{S}_j + \sum_{i \in \mathcal{P}}\hat{S}^2_i
\end{equation}
Measure spin correlation of two different orbital sets $\mathcal{P},\mathcal{Q}$:
\begin{equation}
    \hat{S}_\mathcal{P}\hat{S}_\mathcal{Q} = \sum_{i\in\mathcal{P}}\sum_{j \in \mathcal{Q}} \hat{S}_i\hat{S}_j
\end{equation}
The spin operators $\hat{S}_i^2$ and $\hat{S}_i\hat{S}_j$ can be reduced to:
\begin{equation}
\begin{aligned}
    \hat{S}_i^2 &=  \frac{3}{4}\left(\hat{n}_{i \uparrow}+\hat{n}_{i \downarrow}-2 \hat{n}_{i \uparrow} \hat{n}_{i \downarrow}\right)\\
    \hat{S}_i \hat{S}_j &=  \frac{1}{2}\left(\hat{S}_i^{+} \hat{S}_j^{-}+\hat{S}_i^{-} \hat{S}_j^{+}\right)+\hat{S}_i^z \hat{S}_j^z \\
    & = \frac{1}{2}\left(\hat{c}_{i \uparrow}^{\dagger} \hat{c}_{i \downarrow} \hat{c}_{j \downarrow}^{\dagger} \hat{c}_{j \uparrow}+\hat{c}_{i \downarrow}^{\dagger} \hat{c}_{i \uparrow} \hat{c}_{j \uparrow}^{\dagger} \hat{c}_{j \downarrow}\right)\\
    &\quad \quad +\frac{1}{4}\left(\hat{n}_{i \uparrow} \hat{n}_{j \uparrow}-\hat{n}_{i \uparrow} \hat{n}_{j \downarrow}-\hat{n}_{i \downarrow} \hat{n}_{j \uparrow}+\hat{n}_{i \downarrow} \hat{n}_{j \downarrow}\right)
\end{aligned}
\end{equation}

\begin{figure}
\centerline{\includegraphics[width=\columnwidth]{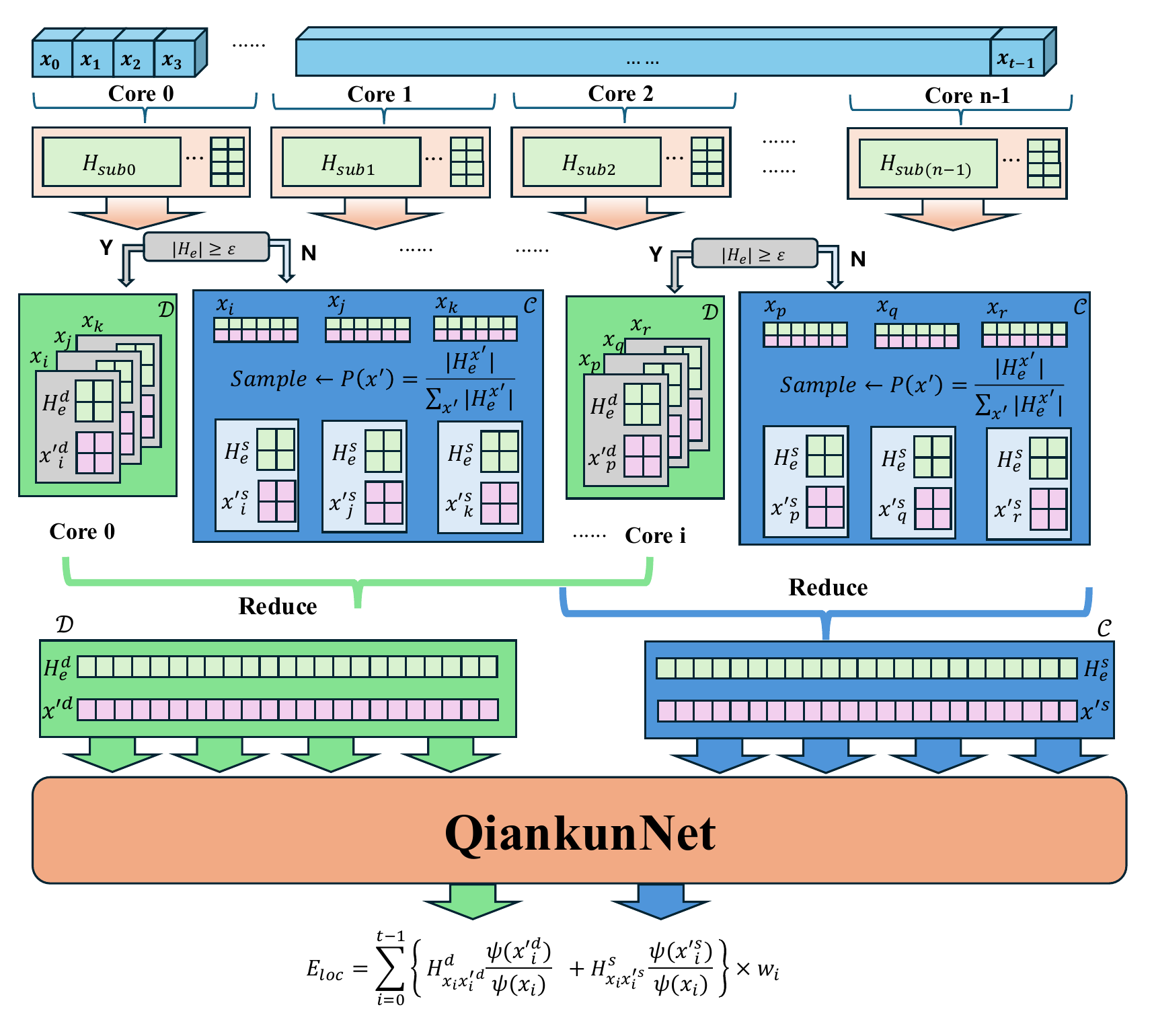}}
\caption{Multi-core local energy computation based on sampled and deterministically obtained coupled states. Each core processes a separate batch of configurations, focusing exclusively on coupling calculations and Hamiltonian matrix elements for its assigned configurations.}
\label{fig:Thread-Level_Parallel}
\end{figure}

\subsection*{Semistochastic Local Energy Computing in Parallel}
The local energy is calculated as: $E_{\text{loc}}=\sum_{\mathbf{x}}H_{\mathbf{xx}^{\prime}}\frac{\psi(\mathbf{x}^{\prime})}{\psi(\mathbf{x})}$, where $\mathbf{x}$ denotes a configuration within the QiankunNet variational space, $\mathbf{x}^{\prime}$ represents the coupled configuration obtained by applying the Hamiltonian to $\mathbf{x}$, and $H_{\mathbf{xx}^{\prime}}$ is the corresponding Hamiltonian matrix element between $\mathbf{x}$ and $\mathbf{x}^{\prime}$. The computational cost of evaluating the local energy primarily arises from two main sources: (1)the first originates from the number of non-zero matrix elements $H_{\mathbf{xx}^{\prime}}$ in the Hamiltonian, which scales as $O(N^{4})$ with system size. This implies that for any given input quantum bitstring $\mathbf{x}$, there can be up to $O(N^{4})$ distinct bitstrings $\mathbf{x}^{\prime}$ such that the matrix element $H_{\mathbf{xx}^{\prime}}$ is non-zero element. This property reflects the strong many-body coupling and highly complex sparse structure of the Hamiltonian in the basis expansion representation, posing significant challenges for numerical matrix operations and sampling strategies. (2)The second time cost comes from evaluating $\psi(\mathbf{x}^{\prime})$. Computing this wavefunction value requires performing a nonlinear mapping from the input bitstring $\mathbf{x}^{\prime}$ to the wavefunction amplitude, a process that typically involves forward propagation through a large number of parameters, attention computation, activation functions across multiple hidden layers, and possibly the evaluation of electron configurations. Its computational complexity far exceeds that of filtering matrix elements under the Slater–Condon rules, thus making it the dominant bottleneck in the overall local energy calculation.

For the first computational hotspot, we efficiently leverage the resources of modern multi-core CPUs by employing thread-level parallelism to accelerate the computation of coupled configurations. For the second hotspot, inspired by Ref.~\cite{wuzibo2025}, we design a semi-stochastic sampling and semi-deterministic coupling scheme to significantly reduce the number of coupled configurations that require explicit evaluation. Fig~\ref{fig:Thread-Level_Parallel} illustrates our efficient many-core parallel architecture for quantum state evolution and local energy computation. The computing module employs a collaborative effort among multiple computing cores, with each core independently processing a assigned batch of configurations, thereby achieving distributed task processing and load balancing.

As shown in the figure, each core maintains a set of specific quantum states $\mathbf{x}$ in its local memory and evaluates the interactions between these states and other configurations through a coupling computation module. QiankunNet determines whether to include a specific transition path into the candidate set by assessing whether the matrix element $H^{\mathbf{x}^{\prime}}_{e}$ exceeds a threshold value $\epsilon$. For those configurations where $|H^{\mathbf{x}^{\prime}}_{e}|>\epsilon$, the QiankunNet algorithm reduces them into the deterministic set $\mathcal{D}$; for those where $|H^{\mathbf{x}^{\prime}}_{e}|<\epsilon$, the QiankunNet temporarily stores them in a candidate set $\mathcal{C}$. Each $|H^{\mathbf{x}^{\prime}}_{e}|$ in $\mathcal{C}$ for a given $x$ is normalized, and sampling is performed according to the probability distribution $P(\mathbf{x}^{\prime})=\frac{|H^{\mathbf{x}^{\prime}}_{e}|}{\sum_{x'}|H^{\mathbf{x}^{\prime}}_{e}|}$, thereby reducing the number of coupling states to be considered. Subsequently, the QiankunNet reduces the local $\mathcal{D}$ and $\mathcal{C}$ from each computing core into global $\mathcal{D}$ and $\mathcal{C}$. Through the integrated QiankunNet network, the resulting wavefunctions $\psi(\mathbf{x}^{\prime})$ are generated. Finally, the total energy expectation value $E$ is efficiently computed via vectorized operations.

\bibliography{refs}

\begin{thebibliography}{10}

\bibitem{Gao2024}
H.~Gao, S.~Imamura, A.~Kasagi, E.~Yoshida, Distributed implementation of full
  configuration interaction for one trillion determinants.
\newblock {\it Journal of Chemical Theory and Computation\/} {\bf 20}, 1185
  (2024). PMID: 38314701.

\bibitem{Perturbation}
T.~Helgaker, P.~J{\o}rgensen, J.~Olsen, {\it {Perturbation Theory}\/} (John
  Wiley and Sons, Ltd, 2000), chap.~14, pp. 724--816.

\bibitem{MP2}
C.~M\o{}ller, M.~S. Plesset, Note on an approximation treatment for
  many-electron systems.
\newblock {\it Phys. Rev.\/} {\bf 46}, 618 (1934).

\bibitem{White1992}
S.~R. White, Density matrix formulation for quantum renormalization groups.
\newblock {\it Phys. Rev. Lett.\/} {\bf 69}, 2863 (1992).

\bibitem{White1993}
S.~R. White, Density-matrix algorithms for quantum renormalization groups.
\newblock {\it Phys. Rev. B\/} {\bf 48}, 10345 (1993).

\bibitem{McMillan1965}
W.~L. McMillan, Ground state of liquid ${\mathrm{he}}^{4}$.
\newblock {\it Phys. Rev.\/} {\bf 138}, A442 (1965).

\bibitem{FoulkesRajagopal2001}
W.~M.~C. Foulkes, L.~Mitas, R.~J. Needs, G.~Rajagopal, Quantum monte carlo
  simulations of solids.
\newblock {\it Rev. Mod. Phys.\/} {\bf 73}, 33 (2001).

\bibitem{AustinLester2012}
B.~M. Austin, D.~Y. Zubarev, W.~A.~J. Lester, Quantum monte carlo and related
  approaches.
\newblock {\it Chem. Rev\/} {\bf 112}, 263 (2012).

\bibitem{MCSCF}
R.~Shepard, {\it The Multiconfiguration Self-Consistent Field Method\/} (John
  Wiley \& Sons, Ltd, 1987), pp. 63--200.

\bibitem{Bartlett2007}
R.~J. Bartlett, M.~Musia\l{}, Coupled-cluster theory in quantum chemistry.
\newblock {\it Rev. Mod. Phys.\/} {\bf 79}, 291 (2007).

\bibitem{CarleoTroyer2017}
G.~Carleo, M.~Troyer, Solving the quantum many-body problem with artificial
  neural networks.
\newblock {\it Science\/} {\bf 355}, 602 (2017).

\bibitem{PfauFoulkes2020}
D.~Pfau, J.~S. Spencer, A.~G. D.~G. Matthews, W.~M.~C. Foulkes, Ab initio
  solution of the many-electron schr\"odinger equation with deep neural
  networks.
\newblock {\it Phys. Rev. Res.\/} {\bf 2}, 033429 (2020).

\bibitem{HermannNoe2020}
J.~Hermann, Z.~Sch{\"a}tzle, F.~No{\'e}, Deep-neural-network solution of the
  electronic schr{\"o}dinger equation.
\newblock {\it Nat. Chem.\/} {\bf 12}, 891 (2020).

\bibitem{vonglehn2022psiformer}
I.~von Glehn, J.~S. Spencer, D.~Pfau, A self-attention ansatz for ab-initio
  quantum chemistry.
\newblock {\it arXiv:2211.13672\/}  (2022).

\bibitem{ChooCarleo2020}
K.~Choo, A.~Mezzacapo, G.~Carleo, Fermionic neural-network states for ab-initio
  electronic structure.
\newblock {\it Nat. Commun.\/} {\bf 11}, 2368 (2020).

\bibitem{BarrettLvovsky2022}
T.~D. Barrett, A.~Malyshev, A.~Lvovsky, Autoregressive neural-network
  wavefunctions for ab initio quantum chemistry.
\newblock {\it Nat. Mach. Intelle.\/} {\bf 4}, 351 (2022).

\bibitem{Zhao2023}
T.~Zhao, J.~Stokes, S.~Veerapaneni, {Scalable neural quantum states
  architecture for quantum chemistry}.
\newblock {\it Machine Learning: Science and Technology\/}  (2023).

\bibitem{PRL2023}
L.~L. Viteritti, R.~Rende, F.~Becca, Transformer variational wave functions for
  frustrated quantum spin systems.
\newblock {\it Phys. Rev. Lett.\/} {\bf 130}, 236401 (2023).

\bibitem{Zhang2023}
Y.-H. Zhang, M.~Di~Ventra, Transformer quantum state: A multipurpose model for
  quantum many-body problems.
\newblock {\it Phys. Rev. B\/} {\bf 107}, 075147 (2023).

\bibitem{Liu2024-1}
A.-J. Liu, B.~K. Clark, Neural network backflow for ab initio quantum
  chemistry.
\newblock {\it Phys. Rev. B\/} {\bf 110}, 115137 (2024).

\bibitem{Wu2023}
Y.~Wu, C.~Guo, Y.~Fan, P.~Zhou, H.~Shang, Nnqs-transformer: An efficient and
  scalable neural network quantum states approach for ab initio quantum
  chemistry.
\newblock {\it Proceedings of the International Conference for High Performance
  Computing, Networking, Storage and Analysis (SC '23)\/} (Association for
  Computing Machinery, New York, NY, USA, 2023).

\bibitem{Fu2024}
L.~Fu, Y.~Wu, H.~Shang, J.~Yang, Transformer-based neural-network quantum state
  method for electronic band structures of real solids.
\newblock {\it Journal of Chemical Theory and Computation\/} {\bf 20}, 6218
  (2024).

\bibitem{Kan2025}
B.~Kan, Y.~Tian, Y.~Wu, Y.~Zhang, H.~Shang, Bridging the gap between
  transformer-based neural networks and tensor networks for quantum chemistry.
\newblock {\it Journal of Chemical Theory and Computation\/} {\bf 21}, 3426
  (2025).

\bibitem{transformer2017}
A.~Vaswani, {\it et~al.\/}, Attention is all you need.
\newblock {\it Advances in neural information processing systems\/} {\bf 30}
  (2017).

\bibitem{radford2018gpt}
A.~Radford, K.~Narasimhan, T.~Salimans, I.~Sutskever, {\it et~al.\/}, Improving
  language understanding by generative pre-training.
\newblock {\it OpenAI blog\/}  (2018).

\bibitem{radford2019gpt2}
A.~Radford, {\it et~al.\/}, Language models are unsupervised multitask
  learners.
\newblock {\it OpenAI blog\/} {\bf 1}, 9 (2019).

\bibitem{Brown2020}
T.~B. Brown, {\it et~al.\/}, Language models are few-shot learners.
\newblock {\it Proceedings of the 34th International Conference on Neural
  Information Processing Systems\/}, NIPS'20 (Curran Associates Inc., Red Hook,
  NY, USA, 2020).

\bibitem{Glehn2022}
I.~von Glehn, J.~S. Spencer, D.~Pfau, A self-attention ansatz for ab-initio
  quantum chemistry.
\newblock {\it ArXiv\/} {\bf abs/2211.13672} (2022).

\bibitem{Feynman1956}
R.~P. Feynman, M.~Cohen, Energy spectrum of the excitations in liquid helium.
\newblock {\it Physical Review\/} {\bf 102}, 1189 (1956).

\bibitem{liu2025NNBF}
A.-J. Liu, B.~K. Clark, Efficient optimization of neural network backflow for
  ab-initio quantum chemistry.
\newblock {\it arXiv preprint arXiv:2502.18843\/}  (2025).

\bibitem{beinert1997iron}
H.~Beinert, R.~H. Holm, E.~Munck, Iron-sulfur clusters: nature's modular,
  multipurpose structures.
\newblock {\it Science\/} {\bf 277}, 653 (1997).

\bibitem{venkateswara2004synthetic}
P.~Venkateswara~Rao, R.~Holm, Synthetic analogues of the active sites of iron-
  sulfur proteins.
\newblock {\it Chemical reviews\/} {\bf 104}, 527 (2004).

\bibitem{sharma2014low}
S.~Sharma, K.~Sivalingam, F.~Neese, G.~K.-L. Chan, Low-energy spectrum of
  iron--sulfur clusters directly from many-particle quantum mechanics.
\newblock {\it Nature chemistry\/} {\bf 6}, 927 (2014).

\bibitem{li2017spin}
Z.~Li, G.~K.-L. Chan, Spin-projected matrix product states: Versatile tool for
  strongly correlated systems.
\newblock {\it Journal of chemical theory and computation\/} {\bf 13}, 2681
  (2017).

\bibitem{yamaguchi1986}
K.~Yamaguchi, Y.~Takahara, T.~Fueno, Ab-initio molecular orbital studies of
  structure and reactivity of transition metal-oxo compounds.
\newblock {\it Applied Quantum Chemistry: Proceedings of the Nobel Laureate
  Symposium on Applied Quantum Chemistry in Honor of G. Herzberg, RS Mulliken,
  K. Fukui, W. Lipscomb, and R. Hoffman, Honolulu, HI, 16--21 December 1984\/}
  (Springer, 1986), pp. 155--184.

\bibitem{schurkus2020theoretical}
H.~Schurkus, D.-T. Chen, H.-P. Cheng, G.~Chan, J.~Stanton, Theoretical
  prediction of magnetic exchange coupling constants from broken-symmetry
  coupled cluster calculations.
\newblock {\it The Journal of Chemical Physics\/} {\bf 152} (2020).

\bibitem{fu2025local}
W.~Fu, {\it et~al.\/}, Local pseudopotential unlocks the true potential of
  neural network-based quantum monte carlo.
\newblock {\it arXiv preprint arXiv:2505.19909\/}  (2025).

\bibitem{Luo2019}
D.~Luo, B.~K. Clark, Backflow transformations via neural networks for quantum
  many-body wave-functions.
\newblock {\it Phys. Rev. Lett.\/} {\bf 122}, 226401 (2019).

\bibitem{kingma2014adam}
D.~P. Kingma, Adam: A method for stochastic optimization.
\newblock {\it arXiv preprint arXiv:1412.6980\/}  (2014).

\bibitem{gu2025solving}
Y.~Gu, {\it et~al.\/}, Solving the hubbard model with neural quantum states.
\newblock {\it arXiv preprint arXiv:2507.02644\/}  (2025).

\bibitem{SR0}
S.~Sorella, Green function monte carlo with stochastic reconfiguration.
\newblock {\it Physical review letters\/} {\bf 80}, 4558 (1998).

\bibitem{SR1}
S.~Sorella, Generalized lanczos algorithm for variational quantum monte carlo.
\newblock {\it Physical Review B\/} {\bf 64}, 024512 (2001).

\bibitem{SR2}
M.~Nightingale, V.~Melik-Alaverdian, Optimization of ground-and excited-state
  wave functions and van der waals clusters.
\newblock {\it Physical review letters\/} {\bf 87}, 043401 (2001).

\bibitem{SR3}
S.~Sorella, M.~Casula, D.~Rocca, Weak binding between two aromatic rings:
  Feeling the van der waals attraction by quantum monte carlo methods.
\newblock {\it The Journal of chemical physics\/} {\bf 127} (2007).

\bibitem{MinSR}
A.~Chen, M.~Heyl, Empowering deep neural quantum states through efficient
  optimization.
\newblock {\it Nature Physics\/} {\bf 20}, 1476 (2024).

\bibitem{MinSR_parallel}
R.~Rende, L.~L. Viteritti, L.~Bardone, F.~Becca, S.~Goldt, A simple linear
  algebra identity to optimize large-scale neural network quantum states.
\newblock {\it Communications Physics\/} {\bf 7}, 260 (2024).

\bibitem{wuzibo2025}
Z.~Wu, B.~Zhang, W.-H. Fang, Z.~Li, Hybrid tensor network and neural network
  quantum states for quantum chemistry (2025).

\end{thebibliography}

\end{document}